\documentclass[12pt]{iopart}
\usepackage{setstack}
\usepackage{graphicx}
\usepackage{amssymb}
\begin{document}
\title{Bounds on the interior geometry and pressure profile of static
  fluid spheres}
\author{Damien Martin, Matt Visser}
\address{School of Mathematical and Computing Sciences, 
Victoria University of Wellington, PO Box 600, Wellington, New Zealand}
\ead{matt.visser@vuw.ac.nz}
\begin{abstract}
  It is a famous result of relativistic stellar structure that (under
  mild technical conditions) a static fluid sphere satisfies the
  Buchdahl--Bondi bound $2M/R\leq8/9$; the surprise here being that
  the bound is not $2M/R\leq1$.  In this article we provide further
  generalizations of this bound by placing a number of constraints on
  the interior geometry (the metric components), on the local
  acceleration due to gravity, on various combinations of the internal
  density and pressure profiles, and on the internal compactness
  $2m(r)/r$ of static fluid spheres.  We do this by adapting the
  standard tool of comparing the generic fluid sphere with a
  Schwarzschild interior geometry of the same mass and radius --- in
  particular we obtain several results for the pressure profile (not
  merely the central pressure) that are considerably more subtle than
  might first be expected.

\vskip 0.50cm
\noindent
  Dated: 10 June 2003; \LaTeX-ed \today
\end{abstract}
\pacs{gr-qc/0306038}
\maketitle
\newtheorem{theorem}{Theorem}
\newtheorem{corollary}{Corollary}
\newtheorem{lemma}{Lemma}
\def\d{{\mathrm{d}}}
\def\implies{\Rightarrow}

\section{Introduction}

A useful pedagogical tool in teaching general relativity is the use of
simple model-independent bounds on stellar structure, or more
generally on compact fluid spheres. In the Newtonian case simple
bounds are nicely summarized by Chandrasekhar, in his book ``Stellar
structure''~\cite{Chandrasekhar}, while in general relativity the most
famous of the model-independent bounds is the Buchdahl--Bondi
limit~\cite{Buchdahl,Bondi} whereby the total mass and radius of any
fluid body is bounded by
\begin{equation}
{2M\over R} <  {8\over9}.
\end{equation}
Note that this is stronger than the obvious bound $2M/R<1$ that would
most naively be expected from the absence of a black hole event
horizon.

Now the Buchdahl--Bondi limit is discussed in many of the standard
textbooks (such as Wald ``General Relativity''~\cite{Wald}, or
Weinberg ``Gravitation and Cosmology''~\cite{Weinberg}) but the
discussion can be greatly enhanced to place limits not only on the
global compactness [$\equiv 2M/R$], but also on additional features
such as:
\begin{itemize}
\item the ``internal compactness''[$\equiv 2m(r)/r$],
\item the locally measured acceleration due to gravity,
\item the metric coefficients of the spacetime,
\item various linear combinations of the pressure and density
  profiles,
\end{itemize} 
at all points throughout the star. The discussion below considerably
extends what we have been able to find in the literature, but because
of its elementary nature is particularly useful when teaching a first
course in general relativity.

Key novel results are a number of bounds on the pressure profile
$p(r)$. These include a relativistic extension of Chandrasekhar's
bound, notable for its simplicity, and a number of constraints
relating the pressure profile to that of the Schwarzschild interior
geometry.  We also discuss the central redshift --- a quantity which
is much less commonly discussed than the surface redshift. While the
surface redshift can be measured directly by looking at optical
spectral lines determining the central redshift would require, for
instance, good spectral information (and theoretical modelling)
regarding neutrino fluxes from the core. Though not a simple task, one
of the payoffs would lie in thereby extracting information regarding
the central pressure from the central redshift --- an inequality on
$z_c$ for general fluid spheres will be one of the results of this
analysis.  As always, bounds of this type are a trade-off between:
\begin{itemize}
\item the ease which with the bound can be stated;
\item the ease which with the bound can be proved;
\item the ease which with the bound can be used;
\item the generality of the bound; and,
\item the strength of the bound.
\end{itemize} 
We have sought a compromise that yields useful physical information
for physically reasonable fluid spheres.

\section{Newtonian Stars}

For a Newtonian star the key structure equation is the equation of
internal equilibrium [we adopt units where $G\equiv 1$]
\begin{equation}
{\d p\over \d r} = - {\rho(r)\; m(r)\over r^2} 
= - {m'(r) \; m(r)\over 4\pi r^4}.
\end{equation}
Chandrasekhar~\cite{Chandrasekhar}, following on and adapting earlier
classical work, uses this to establish several theorems of which one
stands out as the most important for the present purposes:

\begin{theorem}
Provided the average density $\bar\rho(r)= m(r)/[(4\pi/3) r^3]$ is
nonincreasing as one moves outward in the star, then so is the
quantity
\begin{equation}
p(r) + {3\over8\pi} {m(r)^2\over r^4}.
\end{equation}
\end{theorem}

\noindent Proof: 
\begin{eqnarray}
{\d \over \d r}\left[p(r) + {3\over8\pi} {m(r)^2\over r^4}\right] 
&=&
{\d p\over \d r} + {3\over4\pi} {m(r)\; m'(r)\over r^4} 
- {3\over2\pi} {m(r)^2\over r^5}
\\
&=&
{1\over2\pi} {m(r)\over r^4} \left[m'(r) - {3 m(r)\over r}\right]
\\
&=&
{1\over2\pi} {m(r)\over r} {\d \over \d r}\left[{m(r)\over r^3}\right]
\\
&=&
{2\over3} {m(r)\over r} {\d \over \d r}\left[\bar\rho(r)\right] 
\\
&\leq& 0.
\end{eqnarray}

\begin{corollary}
 In a Newtonian star where the average density is nonincreasing outwards 
\begin{equation}
p_c \geq p(r) + {3\over8\pi} {m(r)^2\over r^4} \geq {3\over8\pi} {M^2\over R^4}.
\end{equation}

\end{corollary}
The net result of this theorem is to give robust and useful bounds on
the pressure and density distribution inside a Newtonian star with
only minimal input information. We will seek similar robust and simple
results for general relativistic stars.

\section{TOV Stars}
The key structure equation for a relativistic star is the
Tolman--Oppenheimer--Volkoff [TOV] equation
\begin{equation}
{\d p(r)\over \d r} = - {[\rho(r)+p(r)] \; [m(r)+4\pi p(r) \; r^3] 
\over r^2 [1-2m(r)/r]}.
\end{equation}
The most obvious point to make is that because density, pressure, and
mass are all assumed positive in any reasonable star we now have the
\emph{inequality}
\begin{equation}
{\d p(r)\over \d r} \leq - {\rho(r)\; m(r)\over r^2} 
= - {m'(r) \; m(r)\over 4\pi r^4}.
\end{equation}
Consequently the Newtonian theorem and its corollary still holds in a
relativistic star --- the only modification is in the \emph{proof} of
the theorem; where the equality on going from line 1 to line 2 is now
an inequality. Of course this is a very weak-constraint on the
structure of a relativistic star, and much better constraints are
available. A particularly simple constraint is based on the fact that
for a relativistic star the positivity of pressure implies
\begin{equation}
{\d p(r)\over \d r} \leq - {\rho(r) \; m(r) \over r^2 [1-2m(r)/r]} 
= - {m'(r) \; m(r)\over 4\pi r^4 [1-2m(r)/r]}.
\end{equation}
Note that by dropping both linear and quadratic pressure terms on the
RHS we are implicitly ignoring the ``regeneration of
pressure''~\cite{Harrison}; consequently the bound we derive will be
relatively weak but has the advantage of an elementary proof.

\begin{theorem}
Provided the average density $\bar\rho(r)= m(r)/[(4\pi/3) r^3]$ is
nonincreasing as one moves outward in a relativistic star, then so is
the quantity
\begin{equation}
p(r) - {3\over16\pi} {m(r)\over r^3} \; \ln\left[1-{2m(r)\over r}\right].
\end{equation}
\end{theorem}

\noindent Proof: 
\begin{eqnarray}
&&{\d \over \d r}
\left\{p(r) - {3\over16\pi} {m(r)\over r^3} \; 
\ln\left[1-{2m(r)\over r}\right]\right\}
\\
&&
\qquad\qquad
=
{\d p\over \d r} 
- {3\over16\pi}{\d \over \d r}\left[{m(r)\over r^3}\right] \; 
\ln\left[1-{2m(r)\over r}\right] 
\nonumber\\
&&
\qquad\qquad\qquad\qquad
+ {3\over16\pi} {m(r)\over r^3} {1\over1-2m(r)/ r} 
{\d \over \d r}\left[{2 m(r)\over r}\right]
\\
&&
\qquad\qquad
\leq
{1\over8\pi} {m(r)\over r^4}  {1\over1-2m(r)/ r} 
\left[m'(r) - {3 m(r)\over r}\right] 
\\
&&
\qquad\qquad
=
{1\over8\pi} {m(r)\over r}  {1\over1-2m(r)/ r} 
{\d \over \d r}\left[{m(r)\over r^3}\right]
\\
&&\qquad\qquad
\leq
 0.
\end{eqnarray}

\begin{corollary}
In a relativistic star, provided the average density is nonincreasing
as one moves outward
\begin{equation}
p_c 
\geq 
p(r) - {3\over16\pi} {m(r)\over r^3} \; \ln\left[1-{2m(r)\over r}\right] 
\geq
- {3\over16\pi} {M\over R^3} \; \ln\left[1-{2M\over R}\right].
\end{equation}
\end{corollary}

\noindent Comment:\\
Note that as $2M/R\to 0$ one recovers the Newtonian result, and that
as $2M/R\to 1$ the central pressure is forced to infinity. This
argument is not good enough to reproduce the Buchdahl--Bondi limit,
but it is elementary, and it is sufficiently good to show the runaway
explosion of central pressure for sufficiently compact objects.
Moreover, because in deriving this bound we have not used the
``regeneration of pressure'' effect~\cite{Harrison}, it follows that
the occurrence of unbounded central pressures in general relativity is
not intrinsically a regeneration effect.

To now obtain the Buchdahl--Bondi bound, and its extensions, the best
strategy is to develop a number of comparison theorems that provide
inequalities relating the generic relativistic fluid sphere and the
interior Schwarzschild solution.

\section{Generic interior geometry}

Let us adopt coordinates to write the general interior geometry in the
form
\begin{equation}
\d s^2 = - \zeta(r)^2 \; \d t^2 + 
{\d r^2 \over 1-2m(r)/r} 
+ r^2 \left[ \d \theta^2+\sin^2\theta\; \d\phi^2 \right].
\end{equation}
Equivalently
\begin{equation}
\fl
\d s^2 = 
- \exp\left[-2\int_r^\infty g(\tilde r) \;\d\tilde r\right] \; \d t^2 
+ 
{\d r^2 \over 1-2m(r)/r} 
+ r^2 \left[ \d \theta^2+\sin^2\theta\; \d\phi^2 \right].
\end{equation}
With this choice $g(r)$ is positive for an inward gravitational
attraction and equals the locally measured acceleration due to
gravity. The Einstein equations yield
\begin{eqnarray}
\fl
8\pi \rho = G_{\hat t\hat t} = 2 m'(r)/r^2;
\\
\fl
8\pi p = G_{\hat r\hat r} = 
2 \left\{ 
{g(r) \; \left[1-{2m(r)/ r}\right]\over r} - {m(r)\over r^3} 
\right\};
\\
\fl
8\pi p = G_{\hat \theta\hat \theta} = 
-{ m'(r) [ 1 + r g(r) ]\over r^2}
+{ m(r) [ 1 - r g(r) - 2 r^2 g'(r) - 2 r^2 g(r)^2 ]\over r^3 }
\nonumber\\
\qquad\qquad\qquad
+{ g(r) [ 1 + r g(r) ]\over r}.
\end{eqnarray}
The first of these integrates to
\begin{equation}
m(r) = \int_0^r 4\pi \rho(\tilde r) \; \tilde r^2 \;\d\tilde r,
\end{equation}
which justifies the choices of notation $g_{rr}=1/[1-2m(r)/r]$. The
second equation can be algebraically rearranged to yield
\begin{equation}
g(r) = {m(r)+4\pi p(r) r^3\over r^2 \; [1-2m(r)/r ]}.
\end{equation}
The third equation is quite messy --- however in view of the Bianchi
identities it may be ``traded off'' for the covariant conservation of
stress energy which [for isotropic pressures] takes the rather simple
form
\begin{equation}
p'(r) = - [\rho(r) + p(r) ]\; g(r)
\end{equation}
Eliminating $g(r)$ leads to the TOV equation as previously given.

If we choose to work in terms of $\zeta(r)$ instead of $g(r)$, and we
shall soon see that this is sometimes useful, the only significant
change in appearance [not substance] is in the equation
\begin{equation}
G_{\hat \theta\hat \theta} = 
{\zeta(r)'' [1-2m(r)/r] \over \zeta(r)}
+ {\zeta'(r) [1-m(r)/r-m'(r)]\over r \zeta(r)} 
- {[m(r)/r]'\over r}.
\end{equation}
For the $G_{\hat r\hat r}$ equation there is very little change
\begin{equation}
G_{\hat r\hat r} = 
2 \left\{ 
{\zeta'(r)\over\zeta(r)} \; 
{\left[1-{2m(r)/ r}\right]\over r} - {m(r)\over r^3} 
\right\}.
\end{equation}

\section{Interior Schwarzschild}

The interior Schwarzschild solution is a specific solution of the
Einstein equations corresponding to a star whose density is constant
throughout space. (Despite the common misperception, this does
\emph{not} mean the star is ``incompressible'', see Misner, Thorne,
and Wheeler for details~\cite{MTW}, see esp pp. 609 ff.) If we write
the metric as
\begin{equation}
\d s_*^2 = - \zeta_*(r)^2 \; \d t^2 + 
{\d r^2 \over 1-2m_*(r)/r} + 
r^2 \left[ \d \theta^2+\sin^2\theta\; \d\phi^2 \right],
\end{equation}
then solving the Einstein equations for $\rho=\rho_*$ yields the
standard textbook results~\cite{Wald,Weinberg,MTW}:
\begin{equation}
m_*(r) = {4\pi\over 3} \;\rho_* r^3 = M \; {r^3\over R^3},
\end{equation}
\begin{equation}
\zeta_*(r) = {1\over2} 
\left[ 3 \sqrt{1-{2M/R}} - \sqrt{1-{2m_*(r)/r}} \right],
\end{equation}
and
\begin{equation}
p_*(r) = \rho_* \;
{
\sqrt{1-{2m_*(r)/r}} -  \sqrt{1-{2M/ R}}
\over
3 \sqrt{1-{2M/ R}} - \sqrt{1-{2m_*(r)/r}} 
}.
\end{equation}
Here and henceforth a subscript or superscript star on a quantity
refers to the Schwarzschild interior solution. The central pressure is
\begin{equation}
p_c^* = \rho_* \;
{
1 -  \sqrt{1-{2M/ R}}
\over
3 \sqrt{1-{2M/ R}} - 1
},
\end{equation}
which certainly diverges as $2M/R\to 8/9$; but this is \emph{not yet}
enough to establish the Buchdahl--Bondi bound --- to do so we need to
develop theorems relating the general interior geometry (for an object
with the same mass $M$ and radius $R$) to the specific Schwarzschild
interior geometry.  At the center of the interior Schwarzschild
geometry we also have
\begin{equation}
\zeta_c^* = {1\over2} \left[3\sqrt{1-2M/R}-1\right]
\end{equation}
which is related to the central redshift by
\begin{equation}
z_c^* = {1\over \zeta_c^*} -1 = {3p_c^*\over\rho_*}.
\end{equation} 
As useful algebraic results note that
\begin{equation}
p_*(r) \; \zeta_*(r) = {\rho_*\over2} \;
\left[\sqrt{1-{2m_*(r)/r}} -  \sqrt{1-{2M/ R}}\right],
\end{equation}
\begin{equation}
[\rho_*+p_*(r)] \; \zeta_*(r) = {\rho_*}  \sqrt{1-{2M/R}},
\end{equation}
and
\begin{equation}
[\rho_*+3p_*(r)] \; \zeta_*(r) = {\rho_*}  \sqrt{1-{2m_*(r)/r}}.
\label{e:rho-star}
\end{equation}

\section{Comparison theorems for the spacetime geometry}

Now consider a generic self-gravitating relativistic fluid sphere
\begin{lemma}
If the average density $\bar \rho$ is nonincreasing outwards, then
\begin{equation}
m(r) \geq m_*(r) \equiv M\; {r^3\over R^3}.
\end{equation}
\end{lemma}

\noindent Proof: \\
By definition and hypothesis
\begin{equation}
\bar\rho(r) \equiv {3 m(r)\over 4\pi r^3} \geq \bar\rho(R) 
\equiv {3 M\over 4\pi R^3} \equiv \rho_*.
\end{equation}

\begin{lemma}
  The average density $\bar \rho$ is nonincreasing outwards if and
  only if for all $r$
\begin{equation}
\bar\rho(r) \geq \rho(r).
\end{equation}
\end{lemma}

\noindent Proof: \\
By definition and hypothesis
\begin{equation}
{\d\bar\rho(r)\over\d r} 
= 
{\d\over\d r} \left[ {3 m(r)\over 4\pi r^3} \right]
=
{3\over r} \left[ \rho(r) - \bar\rho(r) \right].
\end{equation}
Note that if the \emph{unaveraged} density $\rho(r)$ is nonincreasing
outwards then certainly $\bar\rho\geq\rho$ and so the average density
$\bar\rho(r)$ is also nonincreasing outwards; the converse however
does not hold. A nonincreasing average density does not necessarily
imply a nonincreasing density and is a genuinely weaker constraint.
From the first lemma it trivially follows that:
\begin{lemma}
If the average density $\bar \rho$ is nonincreasing outwards, then
\begin{equation}
g_{rr}(r) \geq g^*_{rr}(r).
\end{equation}
and
\begin{equation}
m(r)/r \geq m_*(r)/r.
\end{equation}
\end{lemma}
Much more subtle is the following result regarding the $tt$ metric
component:
\begin{theorem}
If the average density $\bar \rho$ is nonincreasing outwards, then 
\begin{equation} 
\zeta(r)
\leq \zeta_*(r).  
\label{e:zeta}
\end{equation} 
Equivalently
\begin{equation}
|g_{tt}(r)| \leq |g^*_{tt}(r)|.
\end{equation}
\end{theorem} 

\noindent Proof: \\
Obtaining this result is somewhat tedious but relatively
straightforward --- an adaptation of the discussion in
Weinberg~\cite{Weinberg} or Wald~\cite{Wald} suffices. With a little
more work, you could also adapt the discussion in the original 1959
Buchdahl paper~\cite{Buchdahl}.  Consider
\begin{eqnarray}
\fl
0 = [G_{\hat r\hat r} - G_{\hat\theta\hat\theta}] \; r^3 \; \zeta(r)
\\
\fl \hphantom{0} =
- r^3 \; \zeta''(r) [1-2m(r)/r] + \zeta'(r) [r^2 + r^2 m'(r) - 3 r m(r)] 
+ \zeta(r) [r m'(r) - 3 m(r)].
\nonumber
\end{eqnarray}
Rearranging
\begin{equation}
{\d\over\d r} \left[ {1\over r} \sqrt{1-{2m(r)\over r}} \; \zeta'(r) \right]
= {\zeta(r)\over \sqrt{1-{2m(r)/ r}}}\; 
  {\d\over\d r} \left[ {m(r)\over r^3} \right ].
\end{equation}
So, as long as the average density is nonincreasing as we go outward
\begin{equation}
{\d\over\d r} \left[ {1\over r} \sqrt{1-{2m(r)\over r}} \; \zeta'(r) \right]
\leq 0.
\label{e:dzeta}
\end{equation}
At the surface of the star $\zeta(r)$ must match smoothly onto the
exterior Schwarzschild geometry, therefore
\begin{equation}
\zeta(R) = \sqrt{1-{2M\over R}}; 
\qquad 
\zeta'(R) = {M\over R^2} {1\over  \sqrt{1-{2M/ R}}} \,.
\end{equation}
Combining (\ref{e:dzeta}) with this boundary condition implies
[$\forall r\in(0,R)$]
\begin{equation}
\zeta'(r) \geq {M\; r\over R^3} \; {1\over \sqrt{1-{2m(r)/ r}}} 
= {m_*(r)\over r^2}.\; {1\over \sqrt{1-{2m(r)/ r}}}
\label{e:dzeta2}
\end{equation}
Integrate from $r$ to $R$, [\emph{not} from 0 to $R$], we have
\begin{equation}
\zeta(R)-\zeta(r) \geq {M\over R^3} 
\int_r^R {\tilde r\over \sqrt{1-{2m(\tilde r)/ \tilde r}}} \; \d\tilde r.
\end{equation}
But in the RHS, $m(r)\geq m_*(r)$, so
\begin{equation}
\fl
\zeta(R)-\zeta(r) \geq {M\over R^3} 
\int_r^R {\tilde r\over \sqrt{1-{2m_*(\tilde r)/ \tilde r}}} \; \d\tilde r
= {M\over R^3} 
\int_r^R {\tilde r \over \sqrt{1-{2M\tilde r^2/R^3}}} \; \d\tilde r.
\end{equation}
This integral is now do-able in closed form, with the result
\begin{equation}
\zeta(R)-\zeta(r) \geq - {1\over 2} 
\left[\sqrt{1-{2M\tilde r^2\over R^3}} \right]_r^R.
\end{equation}
Rearranging, and applying the boundary condition at $R$, we have
\begin{equation}
\zeta(r) \leq {1\over2} 
\left[ 3\sqrt{1-{2M\over R}}  - \sqrt{1-{2M r^2\over R^3}} \right].
\end{equation}
The RHS is now recognizable as $\zeta_*(r)$, the quantity appropriate
to the Schwarzschild interior solution, so the theorem is proved.

\noindent Comment: \\ 
--- Note what has happened here: \emph{Both} of the [physically
nontrivial] metric components, $g_{tt}$ and $g_{rr}$, have been bounded
in terms of what their values would have been for a Schwarzschild
interior solution of the same mass and radius. This now is \emph{more}
than enough to deduce the Buchdahl--Bondi bound. Since $\zeta(r)\leq
\zeta_*(r)$, and since $\zeta_*(r)\to 0$ for some finite positive $r$
whenever $2M/R > 8/9$, we deduce that any star [in which the average
density is nonincreasing outwards] must likewise satisfy the same
bound. More formally:
\begin{corollary}
If the average density $\bar \rho$ is nonincreasing outwards, then
\begin{equation}
{2M\over R} \leq {8\over 9}
\end{equation}
\end{corollary}
\noindent Proof: \\
If the center of the star is to be regular then we must have
\begin{equation}
0 \leq \zeta_c \leq \zeta_c^* = {1\over2} 
\left[ 3\sqrt{1-{2M\over R}}  - 1 \right].
\end{equation}

\noindent Comment: \\ 
--- It might be tempting to conclude that for all $r$
\begin{equation}
\hbox{?`?`?`} \quad {2m(r)\over r} \leq {8\over 9} \qquad ???
\end{equation}
Although true, this conclusion cannot be drawn from the arguments so
far presented.
\\
--- Note also what has \emph{not} happened here. We cannot [at this
stage] deduce \emph{any} bound on the pressures. Indeed it might be
tempting to assert
\begin{equation}
\hbox{?`?`?`} \qquad p(r) \geq p_*(r) \qquad ???
\end{equation}
but such an assertion would actually be \emph{false}. To see this
compare
\begin{equation}
\left. {\d p\over\d r} \right|_R = \rho(R) \; {M\over R^2} \; {1\over 1-2M/R};
\end{equation}
and
\begin{equation}
\left. {\d p_*\over\d r} \right|_R = \rho_* \; {M\over R^2} \; {1\over 1-2M/R}
\geq \left. {\d p\over\d r} \right|_R.
\end{equation}
Thus sufficiently near the surface we must actually have $ p(r) \leq
p_*(r)$; though at the center (as we shall soon see) $p(0) \geq
p_*(0)$. Thus general bounds on the central pressure and pressure
profile are trickier to establish (which is why, regarding the central
pressure, Weinberg resorts to ``it can be shown that'', and Wald
remains silent).

As well as the metric components, there is a similar constraint on the
locally measured acceleration due to gravity detected by static
observers.
\begin{lemma}
If the average density $\bar \rho$ is nonincreasing outwards, then
\begin{equation}
g(r) \geq g^*(r).
\end{equation}
\end{lemma}

\noindent Proof: \\
To see this use  the bound (\ref{e:dzeta2}) to obtain
\[
\zeta'(r) \geq {m_*(r)\over r^2}\; {1\over \sqrt{1-{2m(r)/ r}}}
\geq  {m_*(r)\over r^2}\; {1\over \sqrt{1-{2m_*(r)/ r}}} = \zeta'_*(r),
\]
and then combine this, the definition of $g$, and (\ref{e:zeta}),
to obtain
\begin{equation}
g(r) = {\zeta'(r)\over\zeta(r)} \geq   {\zeta'_*(r)\over\zeta_*(r)} = g^*(r).
\end{equation}

\section{Bounds on the internal compactness $2m(r)/r$}

\begin{theorem}
  If the average density $\bar \rho$ is nonincreasing outwards, then
\begin{equation}
{2 m(r)\over r}  \leq {8\over 9} 
\left[ 
1 + {\sqrt{1+6\pi p(r) r^2}\over 2} -  {1+6\pi p(r) r^2\over 2} 
\right].
\end{equation}
\end{theorem}

\noindent Proof:\\
This particular result is derived, for instance, in Wald, ``General
Relativity''~\cite{Wald}.  To obtain a bound on $2m(r)/r$ modify the
preceding argument by now considering the region $\tilde r\in(0,r)$ with
$r<R$. Then, following the discussion in Wald or using equation
(\ref{e:dzeta}) above we have for $\tilde r \leq r$
\begin{equation}
{1\over \tilde r} \; \sqrt{1-{2m(\tilde r)\over \tilde r}} \; \zeta'(\tilde r) 
\geq
{1\over r} \;\sqrt{1-{2m(r)\over r}} \; \zeta'(r).
\end{equation}
The RHS evaluates to
\begin{equation}
\fl
{1\over r} \sqrt{1-{2m(r)\over r}} \; g(r) \; \zeta(r)
= 
{m(r)+4\pi  p(r) r^3\over r^3  \sqrt{1-{2m(r)/ r}} } \; \zeta(r)
= {4\pi p(r) + m(r)/r^3\over\sqrt{1-{2m(r)/ r}} } \; \zeta(r).
\end{equation}
Thus
\begin{equation}
\zeta'(\tilde r) \geq
\zeta(r) \; {4\pi p(r) + m(r)/r^3\over\sqrt{1-{2m(r)/ r}} }  \;
{\tilde r\over\sqrt{1-2m(\tilde r)/\tilde r}}.
\end{equation}
It is now sufficient to integrate $\tilde r$ from zero to $r$:
\begin{equation}
\zeta(r) - \zeta(0) \geq \zeta(r) 
{4\pi p(r) + m(r)/r^3 \over\sqrt{1-{2m(r)/ r}} }  \; 
\int_0^r {\tilde r\over\sqrt{1-2m(\tilde r)/\tilde r}} \;\d\tilde r.
\end{equation}
That is
\begin{equation}
\zeta(0) \leq \zeta(r) \left\{ 
1 - {4\pi p(r) + m(r)/r^3\over\sqrt{1-{2m(r)/ r}} }  \; 
  \int_0^r {\tilde r\over\sqrt{1-2m(\tilde r)/\tilde r}} \;\d\tilde r.
\right\}
\end{equation}
But because $\zeta(0)$ must be positive in a star whose geometry
remains regular all the way to the center, the quantity in braces must
be positive, and [independent of the unknown value of $\zeta(r)$] we
have the inequality
\begin{equation}
\sqrt{1-{2m(r)\over r}} \geq \left[ 4\pi  p(r) + {m(r)\over r^3}\right] 
\int_0^r {\tilde r\over\sqrt{1-2m(\tilde r)/\tilde r}} \;\d\tilde r.
\end{equation}
But, because the average density is nonincreasing outwards, our first
lemma implies
\begin{equation}
m(\tilde r) \geq m(r) \; {\tilde r^3 \over r^3}.
\end{equation}
This implies
\begin{equation}
\sqrt{1-{2m(r)\over r}} \geq \left[ 4\pi p(r) + {m(r)\over r^3}\right] 
\int_0^r {\tilde r\over\sqrt{1-2m(r)\tilde r^2/r^3}} \;\d\tilde r.
\end{equation}
The integral is now elementary, with the result
\begin{equation}
\sqrt{1-{2m(r)\over r}} \geq \left[ 4\pi p(r) +  {m(r)\over r^3}\right] \;
 {r^3\over 2m(r)} \;
\left\{ 1 - \sqrt{1-{2m(r)\over r}} \right\}.
\end{equation}
Consequently
\begin{equation}
{2m(r)\over r} \sqrt{1-{2m(r)\over r}} \geq 
\left[ 4\pi p(r) r^2 + {m(r)\over r} \right] 
\left\{ 1 - \sqrt{1-{2m(r)\over r}} \right\}.
\end{equation}
Equivalently
\begin{equation}
1-{2m(r)\over r} + \sqrt{1-{2m(r)\over r}} 
\geq 4\pi p(r) r^2 + {m(r)\over r}.
\end{equation}
It is now a matter of tedious but straightforward algebra to rearrange
this inequality into the desired form. Finally, because for positive $x$
one has $\sqrt{1+x}< 1+x$ we deduce:
\begin{corollary}
If the average density $\bar\rho$ is nonincreasing outwards, then
\begin{equation}
{2m_*(r)\over r} \leq {2m(r) \over r} \leq {8\over 9}\,. 
\end{equation}
\end{corollary}

\noindent Proof:\\
Simply use the fact that $p(r)$ is positive inside the star, and the
previously derived lower bound on the compactness. If you prefer you
can write this as
\begin{equation}
2M {r^2\over R^3} \leq {2m(r) \over r} \leq {8\over 9}\,. 
\end{equation}

\noindent Comment:\\
Since $2m(r)/r$ is zero at the center, and equals $2M/R$ at the
surface it is tempting to suppose that $2m(r)/r$ might be monotone
increasing. In general it is not, and many equations of state are
known for which $2m(r)/r$ develops damped oscillatory behaviour as a
function of $r$~\cite{Harrison,Yunes}.

\begin{corollary}
If the average density $\bar\rho$ is nonincreasing outwards, then
\begin{equation}
p(r) \leq {1\over4\pi r^2} \; 
\left\{ 1-{3m(r)\over r} + \sqrt{1-{2m(r)\over r}}\right\}. 
\end{equation}
\end{corollary}

\noindent Comment:\\
This bound is somewhat unusual in that it provides an \emph{upper}
bound on the pressure. It is less useful than one might imagine since
it gives no information about the central pressure.

\section{Comparison theorems for the pressure profile}

General and stringent theorems regarding the central pressure and
pressure profile [rather than metric coefficients and compactness] are
relatively more cumbersome.

With the tools we have at hand, an easy result is this:
\begin{theorem}
If the average density $\bar\rho$ is nonincreasing outwards, then
\begin{equation}
{[{\bar\rho(r)} + 3 \; p(r)] \zeta(r) \over \sqrt{1-2m(r)/r}} 
\geq 
\rho_*
= 
{[{\rho_*} + 3\; p_*(r)] \zeta_*(r)\over\sqrt{1-2m_*(r)/r}} . 
\label{e:strong}
\end{equation}
\end{theorem}

\noindent Proof:\\
First, by the geometric comparison theorems of the previous section,
in particular equation (\ref{e:dzeta2}),
\begin{equation}
\fl
g(r) = {\zeta'(r)\over\zeta(r)} \geq 
{Mr\over R^3} {1\over \sqrt{1-2m(r)/r} \; \zeta(r)}
=
{4\pi \rho_*\; r\over3} {1\over \sqrt{1-2m(r)/r} \; \zeta(r)}.
\end{equation}
Second, by the $\hat r\hat r$  Einstein equation
\begin{equation}
p = {1\over4\pi} \left\{ {g(r)[1-2m(r)/r] \over r} - {m\over r^3} \right\}
= {1\over4\pi}  {g(r)[1-2m(r)/r] \over r} - {\bar\rho(r)\over3}.
\end{equation}
Combine. We have
\begin{equation}
\bar\rho+3p \geq {\rho_* \; \sqrt{1-2m(r)/r}\over \zeta(r)}.
\end{equation}
For the last equality use the properties of the interior Schwarzschild
solution, as embodied in equation~(\ref{e:rho-star}).

\begin{corollary}
If the average density $\bar\rho$ is nonincreasing outwards, then
\begin{equation}
\zeta_c \geq {\rho_*\over \rho_c + 3\;p_c}.
\end{equation}
For the central redshift
\begin{equation}
z_c \leq {\rho_c + 3\;p_c-\rho_*\over \rho_*}.
\end{equation}
\end{corollary}

\begin{corollary}
If the average density $\bar\rho$ is nonincreasing outwards, then
\begin{equation}
{\bar\rho(r)+3 \; p(r) \over \sqrt{1-2m(r)/r} } 
\geq {\rho_* \over\zeta_*(r)} 
= { {\rho_*} + 3\; p_*(r)\over\sqrt{1-2m_*(r)/r}} . 
\end{equation}
\end{corollary}

\noindent Proof:\\
Apply the inequality $\zeta(r)\leq\zeta_*(r)$ to the preceding
theorem.

\begin{corollary}
If the average density $\bar\rho$ is nonincreasing outwards, then
\begin{equation}
{\rho_c}+ 3\;p_c \geq {\rho_*}+ 3\;p_c^*. 
\end{equation}
\end{corollary}

\noindent Comment: \\
While somewhat crude, this bound has the benefit of being both
elementary and guaranteeing that the central value of $\rho+3p$
diverges at or before one reaches the Buchdahl--Bondi limit.  Note
that this does \emph{not} permit us to place a bound on $p_c$ itself,
only on the combination $\rho_c+3p_c$.

The first bound we place on the pressure profile itself is this:
\begin{theorem}
If the average density $\bar \rho$ is nonincreasing outwards, then
\begin{equation}
p(r) \geq {\rho_*\over3} \; 
{
2\sqrt{1-2m(r)/r}+ \sqrt{1-2Mr^2/R^3} - 3 \sqrt{1-2M/R}
\over
3 \sqrt{1-2M/R} -  \sqrt{1-2Mr^2/R^3}
}.
\end{equation}
\end{theorem}

\noindent Proof: \\
Start from the equation $p'=-[\rho+p] g = -
[\rho+p]\zeta'/\zeta$ and re-write it in the form
\begin{equation}
[p \zeta]' = - \rho \zeta'.
\end{equation}
Integrate from $r$ to $R$, then
\begin{equation}
p(r) \; \zeta(r) = 
\int_r^R \rho(\tilde r) \; \zeta'(\tilde r) \; \d\tilde r.
\end{equation}
But because of our earlier inequality on $\zeta'(r)$, embodied in
equation (\ref{e:dzeta2}),
\begin{equation}
p(r) \; \zeta(r) \geq \int_r^R \rho(\tilde r) {M\tilde r\over R^3} 
{1 \over\sqrt{1-2m(\tilde r)/\tilde r}}\d\tilde r
\end{equation}
\begin{equation}
= {M\over R^3}  \int_r^R 
{m'(\tilde r) \over 4\pi \tilde r \sqrt{1-2m(\tilde r)/\tilde r}} \; \d\tilde r
\end{equation}
\begin{equation}
= {\rho_*\over3} \int_r^R 
\left\{ \left[m(\tilde r)\over\tilde r\right]' 
+ {m(\tilde r)\over \tilde r^2 } \right\}
{1\over \sqrt{1-2m(\tilde r)/\tilde r}} \; \d\tilde r
\end{equation}
\begin{equation}
= {\rho_*\over3} 
\left\{ 
- \left.\sqrt{1-2m(\tilde r)/\tilde r}\right|_r^R + 
\int_r^R  
{
{m(\tilde r)/\tilde r^2 }
\over
\sqrt{1-2m(\tilde r)/\tilde r}
} \;\d \tilde r
\right\}.
\end{equation}
Again using the fact that the average density is nonincreasing
outwards
\begin{equation}
p(r) \; \zeta(r) \geq {\rho_*\over3} 
\left\{ 
-\left.\sqrt{1-2m(\tilde r)/\tilde r}\right|_r^R + 
\int_r^R  
{
{M\tilde r/R^3 }
\over
\sqrt{1-2M\tilde r^2/ R^3}
} \;\d \tilde r
\right\}.
\end{equation}
The remaining integral is again elementary, with the consequence
\begin{eqnarray}
\fl
p(r) \; \zeta(r) \geq {\rho_*\over6} 
\left\{ 
-2\sqrt{1-2M/R} + 2\sqrt{1-2m(r)/r} - {\sqrt{1-2M/R}} + \sqrt{1-2Mr^2/R^3}
\right\}
\nonumber
\\
\end{eqnarray}
\begin{equation}
=  {\rho_*\over6} 
\left\{ 
2\sqrt{1-2m(r)/r} + \sqrt{1-2Mr^2/R^3} - 3\sqrt{1-2M/R}
\right\}.
\label{e:pressure-bound}
\end{equation}
Finally, since we have already established $\zeta(r)\leq\zeta_*(r)$,
inserting the explicit form of $\zeta_*(r)$ completes the proof.

\begin{corollary}
If the average density $\bar \rho$ is nonincreasing outwards, then
\begin{equation}
\zeta(r) p(r) -{\rho_*\over3}\sqrt{1-2m(r)/r} 
\geq  \zeta_*(r) p_*(r)  -{\rho_*\over3}\sqrt{1-2m_*(r)/r} 
\end{equation}
\end{corollary}

\noindent Proof:\\
This is a slight strengthening of the previous theorem obtained by not
invoking $\zeta(r)\leq\zeta_*(r)$ at the last step. In this version of
the result everything on the LHS refers to the body we are
investigating and everything on the RHS refers to the interior
Schwarzschild geometry we are using to compare it to.

\begin{corollary}
If the average density $\bar \rho$ is nonincreasing outwards, then
\begin{equation}
p(r) \geq  p_*(r) + {\rho_*\over3} \; 
\left[ {\sqrt{1-2m(r)/r} - \sqrt{1-2m_*(r)/r} \over \zeta_*(r)} \right].
\end{equation}
\end{corollary}

\noindent Proof:\\
This is simply a re-writing of the previous theorem to make it clear
that the generic pressure bound consists of the interior Schwarzschild
result plus a ``correction'' term. The correction term vanishes at
both $R$ and at $0$, and is negative between these locations. Note
that the bound is sharp in the sense that the interior Schwarzschild
solution saturates the bound.

\noindent Comment:\\
Note that to bound the pressure profile $p(r)$ you need information
regarding the local compactness $m(r)/r$, which is difficult to
measure observationally. This contrasts with most of the other bounds
derived in this article which depend on parameters that can be measured
from the stellar exterior.

\noindent Comment:\\
If we take this pressure bound and consider it near the surface of the
star then we find
\begin{equation}
p(R) \geq 0, 
\end{equation}
and
\begin{equation}
\left.{\d p\over\d r}\right|_R \geq - \rho(R)\; {M\over R^2 (1-2M/R)}.
\end{equation}
But, because of the boundary conditions at the surface of the star,
these bounds are actually \emph{saturated} by any fluid sphere, and
cannot be improved.

\begin{corollary}
If the average density $\bar \rho$ is nonincreasing outwards, then
\begin{equation}
p_c \geq  p_c^*.
\end{equation}
\end{corollary}

\noindent Comment:\\
Thus the interior Schwarzschild solution really does provide a lower
bound on the central density even if it cannot be used [in
``uncorrected'' form] elsewhere within the star. Consequently, in the
Buchdahl--Bondi limit $2M/R\to8/9$ not only does $\zeta_c\to 0$, but
we also have $p_c\to \infty$ [as intuitively expected].

We now consider several additional bounds that are derived in slightly
different fashion.
\begin{theorem}
  If the density $\rho$ [not the average density] is nonincreasing
  outwards, then in terms of the surface density $\rho_s$:
\begin{equation}
p(r) + \rho(r) \geq \rho_s\;{\sqrt{1-2M/R}\over\zeta(r)} 
\geq  \rho_s\;{\sqrt{1-2M/R}\over\zeta_*(r)}.
\end{equation}
\end{theorem}

\noindent Proof: 
\\
Start from the equation $p'=-[\rho+p] \; g = - [\rho+p]\zeta'/\zeta$
and re-write it in the form
\begin{equation}
[(p+\rho) \; \zeta]' = + \rho'  \zeta \leq 0
\end{equation}
Integrate from $r$ to $R$, then
\begin{equation}
\rho_s\;\sqrt{1-2M/R} - [p(r) + \rho(r)]\;\zeta(r)\leq 0.
\end{equation}
Rearrange, and use the previous inequality for $\zeta(r)$.

\noindent Comment: \\
This is a relatively weak bound because it involves the surface
density $\rho_s$. On the other hand, the RHS of the bound is now
expressed in terms of quantities that are in principle accessible via
external observation.

\begin{corollary}
If the density $\rho$ is nonincreasing outwards, then
\begin{equation}
p_c + \rho_c  \geq \rho_s\;{\sqrt{1-2M/R} \over\zeta^*_c} 
=  
\rho_s \; {2\sqrt{1-2M/R}\over 3\sqrt{1-2M/R} - 1} 
.
\end{equation}
\end{corollary}
The theorem above can be considerably strengthened by noting that
$p'=-[\rho+p] g = - [\rho+p]\zeta'/\zeta$ implies the exact general
result
\begin{equation}
\zeta(r) = \zeta_c
\exp\left[ -
\int_0^r {\d p/\d r\over \rho(\tilde r)+ p(\tilde r)} \; \d \tilde r
\right].
\end{equation}
But, assuming the density is nonincreasing outwards, we have $\rho_c
\geq \rho(\tilde r) \geq \rho(r)$, whence
\begin{equation}
\zeta(r) \geq \zeta_c
\exp\left[ -
\int_0^r {\d p/\d r\over \rho_c + p(\tilde r)} \; \d \tilde r
\right] = \zeta_c \; {\rho_c + p_c\over \rho_c + p(r)}.
\end{equation}
Similarly
\begin{equation}
\zeta(r) \leq \zeta_c
\exp\left[ -
\int_0^r {\d p/\d r\over \rho(r)+ p(\tilde r)} \; \d \tilde r
\right] = \zeta_c \; {\rho(r) + p_c\over \rho(r) + p(r)}.
\end{equation}
We summarize this as a theorem:
\begin{theorem}
  If the density $\rho$ [not the average density] is nonincreasing
  outwards, then 
\begin{equation}
\zeta(r) \geq \zeta_c \; {\rho_c + p_c\over \rho_c + p(r)};
\qquad
\hbox{and}
\qquad
\zeta(r) \leq \zeta_c \; {\rho(r) + p_c\over \rho(r) + p(r)}.
\end{equation}
\end{theorem}
If we now evaluate this at the surface of the body:
\begin{corollary}
  If the density $\rho$ [not the average density] is nonincreasing
  outwards, then
\begin{equation}
\sqrt{1-2M/R}  \geq \zeta_c \; {\rho_c+p_c\over\rho_c};
\quad
\hbox{and}
\quad
\sqrt{1-2M/R}  \leq \zeta_c \; {\rho_s+p_c\over\rho_s}.  
\end{equation}
\end{corollary}
A closely related result is obtained by similar bounds applied to the
exact general result
\begin{equation}
\zeta(r) = \sqrt{1-2M/R} \;
\exp\left[ 
\int_r^R {\d p/\d r\over \rho(\tilde r)+ p(\tilde r)} \; \d \tilde r
\right].
\end{equation}
Without repeating the details we simply quote the results.
\begin{theorem}
  If the density $\rho$ [not the average density] is nonincreasing
  outwards, then 
\begin{equation}
\zeta(r) \geq \sqrt{1-2M/R}  \;\; {\rho_s\over \rho_s + p(r)};
\end{equation}
and
\begin{equation}
\zeta(r) \leq  \sqrt{1-2M/R} \;\; {\rho(r) \over \rho(r) + p(r)}.
\end{equation}
\end{theorem}
These constraints bound the metric components in terms of the pressure
and density profiles, and vice versa. A qualitatively different type
of constraint is provided by the following result.
\begin{theorem}
  If the sum of the pressure and density is positive
\begin{equation}
\sqrt{1-2m(r)/r} \geq \zeta(r) \geq \zeta_c \; \sqrt{1-2m(r)/r}
\end{equation}
\end{theorem}

\noindent 
Proof: Consider
\begin{equation}
{\d\over\d r} \left[ {\zeta(r)\over  \sqrt{1-2m(r)/r} }\right] 
=  {\zeta(r)\over  \sqrt{1-2m(r)/r}} \; 
\left[ g + {[m(r)/r]'\over1-2m(r)/r}\right].
\end{equation}
\begin{equation}
= {\zeta(r)\over  [1-2m(r)/r]^{3/2} } 
\left[ {m + 4\pi p r^3\over r^2} + {4\pi\rho r^2\over r} - {m\over r^2} \right].
\end{equation}
\begin{equation}
= 4\pi r\;{\zeta(r)\over [1-2m(r)/r]^{3/2} } \; [\rho + p] \geq 0.
\label{e:nec}
\end{equation}
Therefore
\begin{equation}
\zeta_c  \leq {\zeta(r)\over  \sqrt{1-2m(r)/r}} \leq 1.
\end{equation}

\noindent Comment: \\
Note that this bound makes no reference to the interior
Schwarzschild solution for comparison purposes. 

\noindent Comment: \\
The condition $\rho+p\geq0$ is the so-called null energy condition
[NEC] --- it is certainly satisfied in bulk nuclear matter, though
there are exotic situations (typically quantum) in which the NEC is
violated~\cite{twilight}. For the purposes of the present article the
NEC is automatically assumed in view of our basic hypotheses that
pressure and density are both individually positive within the fluid
body.

\begin{corollary}
  If the sum of the pressure and density is positive
\begin{equation}
\zeta_c^2 \leq |g_{tt}(r)| \; g_{rr}(r)  \leq 1.
\end{equation}
\end{corollary}
We can now tighten this result by keeping track of the terms we know
to be positive a little bit longer. Re-write equation (\ref{e:nec})
in the form
\begin{equation}
{\sqrt{1-2m(r)/r}\over\zeta(r)} \;
{\d\over\d r} \left[ {\zeta(r)\over  \sqrt{1-2m(r)/r} }\right] 
= 
4\pi r\;{\rho + p\over1-2m(r)/r }.
\end{equation}
Whence
\begin{equation}
{\zeta(r)\over  \sqrt{1-2m(r)/r}} = 
\exp\left[ - \int_r^R 4\pi \tilde r\;\;{\rho(\tilde r) + p(\tilde r)
\over1-2m(\tilde r)/\tilde r } \; \d\tilde r \right].
\end{equation}
To re-write the RHS, we use the TOV equation written in the form
\begin{equation}
{\d p\over\d r} = - {4\pi r(\rho+p)(\bar\rho+3p)\over3(1-2m(r)/r)}
\end{equation}
to obtain 
\begin{equation}
{\zeta(r)\over  \sqrt{1-2m(r)/r}} = 
\exp\left[  
\int_r^R {3 \; \d p/\d r\over \bar\rho(\tilde r)+ 3p(\tilde r)} \; \d \tilde r
\right].
\end{equation}
But now if we use our standard assumption that the average density is
nonincreasing outwards we have first, $\bar\rho\geq\rho_*$. Therefore
$1/(\bar\rho+3p) \leq 1/(\rho_*+3p)$, and because $\d p/\d r$ is
negative, $p'/(\bar\rho+3p) \geq p'/(\rho_*+3p)$. Then
\begin{equation}
{\zeta(r)\over  \sqrt{1-2m(r)/r}} 
\geq
\exp\left[ \int_r^R {3 \; \d p/\d r\over \rho_*+ 3p(\tilde r)} \; \d \tilde r
\right] 
= {\rho_*\over\rho_*+3p(r)}.
\end{equation}
Conversely, since $\tilde r \in (r,R)$, we have $\bar\rho(\tilde r)
\leq \bar\rho(r)$. Therefore $1/[\bar\rho(\tilde r)+3p(\tilde r)] \geq
1/[\bar\rho(r)+3p(\tilde r)]$, and because $\d p/\d r$ is negative,
$p'/[\bar\rho(\tilde r) +3p(\tilde r)] \leq p'/[\bar\rho(r)+3p(\tilde
r)]$. Then performing the integration yields
\begin{equation}
{\zeta(r)\over  \sqrt{1-2m(r)/r}} 
\leq
\exp\left[ \int_r^R {3 \; \d p/\d r\over \bar\rho(r)+ 3p(\tilde r)} \; \d \tilde r
\right] 
= {\bar\rho(r)\over\bar\rho(r)+3p(r)},
\end{equation}
and we have our penultimate theorem:
\begin{theorem} 
If the average density $\bar \rho$ is nonincreasing outwards, then
\begin{equation}
{\zeta(r)[\rho_*+3p(r)]\over  \sqrt{1-2m(r)/r}}  \geq \rho_*;
\qquad
\hbox{and}
\qquad
{\zeta(r)[\bar\rho(r)+3p(r)]\over  \sqrt{1-2m(r)/r}}  \leq \bar\rho(r).
\end{equation}
\end{theorem}
Note that the RHS of the first inequality is independent of location
within the gravitating body.  Note that this result is definitely more
restrictive than equation (\ref{e:strong}), which was obtained by
somewhat simpler techniques.  This theorem has several interesting
corollaries, easily derived using the comparison bounds on $\zeta(r)$
and $m(r)$. It is easiest to first write
\begin{corollary}
If the average density $\bar \rho$ is nonincreasing outwards, then
\begin{equation}
{\zeta(r)\; [\rho_*+3p(r)]\over  \sqrt{1-2m(r)/r}}  
\geq \rho_*
=
{\zeta_*(r)\; [\rho_*+3p_*(r)]\over  \sqrt{1-2m_*(r)/r}}.
\end{equation}
\end{corollary}
Then, since $\zeta(r)\leq\zeta_*(r)$:
\begin{corollary}
If the average density $\bar \rho$ is nonincreasing outwards, then
\begin{equation}
{\rho_*+3p(r)\over  \sqrt{1-2m(r)/r}}  
\geq
{\rho_*+3p_*(r)\over  \sqrt{1-2m_*(r)/r}}.
\end{equation}
\end{corollary}
Similarly, using the fact that $m(r)\geq m_*(r)$:
\begin{corollary}
If the average density $\bar \rho$ is nonincreasing outwards, then
\begin{equation}
{p(r)\over  \sqrt{1-2m(r)/r}}  
\geq
{p_*(r)\over  \sqrt{1-2m_*(r)/r}}.
\end{equation}
\end{corollary}
Though the derivation was somewhat complex and lengthy, this last
result is by far the cleanest looking bound we have derived on the
pressure profile.

Finally, for the central redshift we have:
\begin{corollary}
If the average density $\bar \rho$ is nonincreasing outwards, then
\begin{equation}
\zeta_c \geq {\rho_*\over\rho_*+3p_c};
\qquad
\hbox{and}
\qquad
\zeta_c \leq {\rho_c\over\rho_c+3p_c};  
\end{equation}
so that
\begin{equation}
{3p_c\over\rho_c} \leq z_c \leq {3p_c\over\rho_*},
\end{equation}
and
\begin{equation}
{\rho_* \; z_c\over 3} \leq p_c \leq {\rho_c \; z_c\over 3}.
\end{equation}
\end{corollary}
Finally, for our last theorem we start from the exact result
\begin{equation}
{\zeta(r)\over  \sqrt{1-2m(r)/r}} = \zeta_c \;
\exp\left[  -
\int_0^r {3 \; \d p/\d r\over \bar\rho(\tilde r)+ 3p(\tilde r)} \; \d \tilde r
\right],
\end{equation}
and again bound $\rho_c \geq \bar\rho(\tilde r) \geq \bar\rho(r)$.
Without repeating the details is is now easy to see that:
\begin{theorem}
If the average density $\bar \rho$ is nonincreasing outwards, then
\begin{equation}
{\zeta(r)\over  \sqrt{1-2m(r)/r}} \geq 
\zeta_c \; {\rho_c+3p_c\over\rho_c+3p(r)},
\end{equation}
and
\begin{equation}
{\zeta(r)\over  \sqrt{1-2m(r)/r}} \leq 
\zeta_c \;{\bar\rho(r) +3p_c\over\bar\rho(r)+3p(r)}.
\end{equation}
\end{theorem}

\section{Conclusion}

We have extended classic work by Buchdahl~\cite{Buchdahl} and
Bondi~\cite{Bondi} to derive several model-independent limits on
stellar structure that go beyond what is commonly found in the
literature. (Several important papers are those of
Kovetz~\cite{Kovetz}, Islam~\cite{Islam}, and
Forrester~\cite{Forrester}; despite the wide variety of inequalities
derived in those papers the constraints on the pressure profile
derived above seem to be new.)  Many of the theorems in this present
article are of the form ``If the average density is nonincreasing
outwards, then generic stellar structure is in many ways bounded by
the interior Schwarzschild solution of the same mass and radius''.
While this is not unexpected, classic results typically refer to
global aspects such as the overall compactness [$=2M/R$] or the
central pressure $p_c$ --- in this article we have seen how to extend
this to local information regarding the metric coefficients and, local
compactness, and pressure profile at all points interior to the star.
The bounds we derive are sharp in the sense that they are saturated by
the interior Schwarzschild geometry, so that in this sense they are
the ``best possible''.

Interest in these issues arose while teaching a course in general
relativity and noting that while many textbooks discuss the
Buchdahl--Bondi bound, most did not do so by name, and no extant
textbook [or, so far as we are aware, the technical literature] covers
all the inequalities derived herein.

\noindent
Further issues that might bear looking at:
\\
--- Are there any comparable bounds under more stringent hypotheses?
\\
(For Newtonian stars making the stronger assumption that $\rho$ itself
is nonincreasing yields somewhat better bounds on the pressure
profile~\cite{Chandrasekhar}. We have used this hypothesis for some of
our theorems but suspect that even better bounds may be possible.)
\\
--- Are there any comparable nice \emph{upper} bounds on the pressure?
\\
(We have already seen one upper bound on the pressure profile which
unfortunately never gives any information about the central pressure.
For some progress on this question see Kovetz~\cite{Kovetz}.)
\\
--- Are there similar bounds for solid objects such as planets?
\\
(This leads into the subject of ``anisotropic stars'' for which some
of the best bounds available seem to be those of Guven and
O'Murchadha. See~\cite{Guven} and references therein.)
\\
--- Are there similar bounds for a liquid star enclosed in a solid
crust?
\\
(This would require some hybrid techniques --- a fluid interior with
an anisotropic crust.)

In closing, we reiterate that while the techniques used to derive
these bounds are ``elementary'' in the sense that they can be
explained in a introductory course on general relativity, the results
are rather non-trivial.

\appendix

\ack

This Research was supported by the Marsden Fund administered by the
Royal Society of New Zealand.

\section*{References}



\begin{thebibliography}{99}

\bibitem{Chandrasekhar} 
S.~Chandrasekhar,
``Principles of stellar dynamics'',
(University of Chicago Press, Chicago, 1942).

\bibitem{Buchdahl} 
H.~A.~Buchdahl,
``General Relativistic fluid spheres'',
Phys.~Rev.~{\bf 116} (1959) 1027--1034;
\\
``General Relativistic fluid spheres II: 
general inequalities for regular spheres'',
Ap. J. {\bf 146} (1966) 275--281.


\bibitem{Bondi}
H.~Bondi,
``Massive spheres in general relativity'',
Mon.~Not.~Roy.~Astron.~Soc.  {\bf 282} (1964) 303--317.

\bibitem{Wald}
R.~M.~Wald,
``General Relativity'',
(University of Chicago Press, Chicago, 1984).

\bibitem{Weinberg}
S.~Weinberg,
``Gravitation and cosmology: 
principles and applications of the general theory of relativity'',
(Wiley, New York, 1972).

\bibitem{Harrison}
B.~K.~Harrison, K.~S.~Thorne, M.~Wanako and J.~A.~Wheeler,
``Gravitational theory and gravitational collapse'',
(University of Chicago Press, Chicago, 1965).

\bibitem{MTW}
C.~W.~Misner, K.~S.~Thorne and J.~A.~Wheeler,
``Gravitation'',
(W.~H.~Freeman, San Francisco, 1973).

\bibitem{Yunes} 
M.~Visser and N.~Yunes,
``Power laws, scale invariance, and generalized Frobenius series: 
Applications to Newtonian and TOV stars near criticality'',
arXiv:gr-qc/0211001.

\bibitem{twilight} 
C.~Barcelo and M.~Visser,
``Twilight for the energy conditions?,''
Int.~J.~Mod.~Phys.~D {\bf 11} (2002) 1553
[arXiv:gr-qc/0205066].

\bibitem{Kovetz}
A. Kovetz,
``Some extremal properties of massive spheres in general relativity'',
Ap. J. {\bf 154} (1968) 241--250;
\\
``Minimal and maximal values of the central pressure and temperature 
in convectively stable stars'',
Mon.~Not.~Roy.~Astron.~Soc.  {\bf 144} (1969) 459-460.


\bibitem{Islam}
J.~N.~Islam,
``Some general relativistic inequalities for a star in in 
hydrostatic equilibrium'',
Mon.~Not.~Roy.~Astron.~Soc.  {\bf 145} (1969) 21--29;
\\
``Some general relativistic inequalities for a star in in 
hydrostatic equilibrium II'',
Mon.~Not.~Roy.~Astron.~Soc.  {\bf 147} (1970) 377--386.

\bibitem{Forrester}
D.~A.~Forrester,
``Some relativistic integral theorems'',
Mon.~Not.~Roy.~Astron.~Soc.  {\bf 51} (1970) 149-156.

\bibitem{Guven} 
J.~Guven and N.~O'Murchadha,
``Bounds on 2m/R for static spherical objects'',
Phys.~Rev.~D {\bf 60} (1999) 084020
[arXiv:gr-qc/9903067].




\end{thebibliography}
\end{document}